\documentclass[prb,twocolumn,twoside,floatfix,showpacs]{revtex4}
\usepackage{graphicx}
\usepackage{graphicx,amsfonts}
\usepackage{epsfig,amsmath}
\usepackage{verbatim}

\begin{document}

\title{Disordered free fermions and the Cardy Ostlund fixed line at low
  temperature} 
\author{Pierre Le Doussal}
\affiliation{LPTENS CNRS UMR 8549 24, Rue Lhomond 75231 Paris
Cedex 05, France}
\author{Gr\'egory Schehr}
\affiliation{Theoretische Physik Universit\"at des Saarlandes
66041 Saarbr\"ucken Germany}
\affiliation{Laboratoire de Physique Th\'eorique (UMR du
  CNRS 8627), Universit\'e de Paris-Sud, 91405 Orsay Cedex,
  France} 
\date{\today}

\begin{abstract}
Using functional RG, we reexamine the glass phase of the 2D
random-field Sine Gordon model. It is described by a line of fixed
points (FP) with a super-roughening amplitude
$\overline{(u(0)-u(r))^2} \sim A(T) \ln^2 r $ as temperature $T$ is
varied. A speculation is that this line is identical to the one
found in disordered free-fermion models via exact results from
``nearly conformal'' field theory. This however predicts $A(T=0)=0$,
contradicting numerics. We point out that this result may be related
to failure of dimensional reduction, and that a functional RG method
incorporating higher harmonics and non-analytic operators predicts a
non-zero $A(T=0)$ which compares reasonably with numerics.
\end{abstract}
\maketitle

\section{introduction}

\subsection{Overview}

Few analytical methods can handle systems with quenched disorder
beyond mean field and phenomenological
arguments, and each seems adapted only to a specific
class of models. One example is the functional RG (FRG)
\cite{DSFisher1986}
used to study elastic objects in a random potential
\cite{reviews_pinning,bg_long}
or continuous (e.g. XY) spin models in a random field
\cite{feldman_rf}.
When pure, these systems possess internal order and one studies
the case where the (weak) quenched disorder produces only elastic
deformations (no topological defects) parameterized
by a displacement field $u(x)$ away from perfect order $u=0$.
They exhibit glass physics and pinning,
i.e. thermal (and quantum) fluctuations are small and energy
minimization yields multiple metastable states. The FRG
handles this by summing an infinite set of
operators, all relevant for internal dimension $d<d_{uc}$,
usually $d_{uc}=4$. The coupling constant is a function,
$R(u)$, which represents the correlator of the coarse-grained
disorder. FRG fixed points $R^*(u)$ were obtained in a $d=d_{uc}-\epsilon$
expansion for various disorder classes \cite{DSFisher1986,bg_long}
and recently it was shown how to directly measure
them in numerics \cite{pierre_rofu}.

Disordered systems in $d=2$ can sometimes be studied using
methods inspired by conformal field theory (CFT). Exact results were obtained
for free fermions with disorder of various symmetries
using bozonisation and supersymmetry \cite{ludwig,mudry}.
The physics there is localization, and in recently much studied symmetry classes,
de-localization near a point where the density of state (DOS) is singular
(e.g. bipartite lattices leading to Dirac fermions). There, sample (and local DOS)
fluctuations can become broad leading to freezing transitions in the DOS
dynamical exponent $z$, $\rho(E)\sim E^{-1+2/z}$, as found for Dirac
fermions with random vector potential \cite{baruchpld}. Being equivalent
to a (dilute) limit of a random gauge XY spin model it exhibits an infinite set of relevant
operators \cite{nattermann95,chamon} which can be handled using {\it another} functional RG method
\cite{dcpld}
based on the KPP-Fisher equation (hence noted here KPP-FRG).
Freezing transitions are expected to impact a wider class of models,
e.g. the time-reversal, particle-hole symmetric class,
where the $\rho(E) \sim E^{-1} e^{-c (\ln E)^\mu}$ Gade-Wegner
singularity ($\mu=1/2$) was argued to become $\mu=2/3$ \cite{motrunich}.
This was first found using KPP-FRG on the dual spin version
(the Cardy-Ostlund model with dilute vortices) \cite{tgphysica} and, recently,
within the fermion model \cite{mudry,fukui} and the sigma model \cite{danna}.
These emerging 2d field theories still need testing
on whether they capture all (nonperturbative) freezing (and glassy)
physics, for which the spin models can provide insight.

\subsection{Model}

In this paper we focus on the simpler 2D random field XY (Cardy-Ostlund) model
{\it without} vortices \cite{co} . Being the bosonized form of a fermion model, exact results
were claimed for this model using CFT inspired methods \cite{ludwig,mudry}.
Since it is also a (periodic) pinning problem, we can also use FRG and test
the strong disorder regime, expected to be correctly described
already to one loop. The model, also called random
phase Sine Gordon model (refered to here simply as CO model)
also describes a periodic object (such as an array of flux lines)
at equilibrium on a 2d disordered substrate \cite{natter1, bg_long}
at temperature $T$, and of partition function $Z = \int {\cal D}u e^{- {\cal S_{\rm
      CO}}[u]}$ with ${\cal S_{\rm CO}}[u]= H_{\rm CO}[u]/T$ and:
\begin{equation}
H_{\rm CO}[u] = \int d^2 r [ \frac{\kappa}{2}(\nabla_r u)^2 - h \cdot \nabla_r u
- {\rm Re}(\xi
e^{i u} ) ]  \label{def_co}
\end{equation}
where $u(r) \in ]-\infty, +\infty[$ is an XY phase (excluded
vortices), $\kappa$ the elastic coefficient and $\xi(r)$ a (complex)
quenched random field with random phase $\overline{\xi(r)\xi^*(r')}
= 2 g_0 \delta^{2}(r-r')$. The additional random field $h(r)$ is
generated under coarse-graining with $\overline{h^i(r) h^j(r')}=
\sigma \delta_{ij} \delta^{2}(r-r')$. Disorder being short ranged,
the model (\ref{def_co}), together with some of its discrete
versions, exhibits the STS symmetry \cite{sts,fisherhwa}, i.e.
statistical invariance of the non-linear part under $u(r) \to u(r) +
v(r)$. The model (\ref{def_co}) exhibits a super-rough phase for $T
< T_g = 4\pi \kappa$ with anomalous growth of the 2-point
correlation \cite{co, toner, fisherhwa}:
\begin{eqnarray}
B(r) = \overline{\langle(u(r) - u(0))^2\rangle} = A(T) (\ln{r})^2 +
{\cal O}(\ln{r}) \label{roughness}
\end{eqnarray}
where $\langle ... \rangle$ denotes thermal averages. This is in
contrast with the usual logarithmic roughness for $T>T_g$,
disorder being irrelevant, {\it i.e.} $A(T>T_g) = 0$. 
The STS symmetry implies that $T/T_g$
is uncorrected by disorder and does not flow under RG.
This ''marginal glass'' below $T_g$ is thus described
at large scale by {\it a line a fixed points} (FP), indexed by $T/T_g$
(which can be identified from the large $r$ thermal
connected correlation $\overline{\langle (u(r) - u(0))^2\rangle_c}
\sim 4T/T_g \ln r$). The line of FP can be reached near $T_g$ by perturbative
one loop RG which predicts \cite{carpentier_triang}:
\begin{equation}
A(T) = 2\tau^2 + {\cal O}(\tau^3) \quad \text{for small} \quad  \tau = (T_g-T)/T_g \ll 1
\end{equation}
in agreement with precise numerics \cite{simu_tg}.
At low temperature, only numerical results are available
at present: it was concluded from exact ground state determinations
\cite{rieger, middleton} for a closely related solid on solid model
that super-roughening holds at $T=0$, i.e. $A(T=0) \approx .5 >0$
(see below).

\section{relations to disordered fermions}

\subsection{General framework and bosonization}

A recent study \cite{ludwig} of a free fermion model opened hope
that $A(T)$ could be obtained exactly. It has been
known that weak disorder on the CO model at any $T$
can be equivalently studied by adding quenched (dynamical) disorder to a (euclidean) $d=1+1$ {\it
interacting} fermion model \cite{tgpldrsb} . Starting with the CO model {\it
without disorder}, i.e. $h,\xi=0$ in (\ref{def_co}), one defines
$r=(x,\tau)$ where $\tau \in [0,\beta \hbar]$ is imaginary time, and
$u(r)=2 \phi(r)$. Then the bosonic action ${\cal
S}_{b}[\phi]={\cal S}_{\rm CO}[u=2 \phi]$, where one substitutes
$\frac{\kappa}{T} \nabla_r^2 \to \frac{1}{2 \pi K} (\frac{1}{v}
\partial_\tau^2 + v \partial_x^2)$ in (\ref{def_co}), is the euclidean action (with $Z=\int
{\cal D}\phi e^{- {\cal
    S}_b[\phi]}$) of the bozonized form of the fermion model (with
$Z=\int {\cal D}\psi^\dagger {\cal D}\psi e^{- {\cal
    S}_f[\psi,\psi^\dagger]}$) of
euclidean action ${\cal S}_f = S_0 + S_{int}$ with:
\begin{eqnarray} \label{eq:fermions-periodic}
&& S_0 = \int_{x \tau} [ \psi_R^\dagger \partial_\tau \psi_R + \psi_L^\dagger \partial_\tau \psi_L ] + \frac{1}{\hbar} \int_\tau H_0 \\
&& H_0 = \int_x  \hbar v_F : \psi_L^\dagger (i \partial_x -k_F) \psi_L -
\psi_R^\dagger (i \partial_x + k_F) \psi_R: \nonumber \\
&& S_{int} = \frac{1}{\hbar} \int_{x \tau} \frac{g_4}{2} (\rho_R
\rho_R + \rho_L \rho_L) + g_2 \rho_R \rho_L
\end{eqnarray}
where $\int_{x\tau}=\int dx \int_0^{\beta \hbar} d\tau$,
$\psi=\psi_L +\psi_R$, $\rho=\psi^\dagger \psi$ is the fermion
density and $g_{2,4}$ are (spinless) fermion interaction parameters
with $\rho_s=\psi_s^\dagger \psi_s$, $s=L,R$. Using
bosonization:
\begin{eqnarray}
&& \psi_s=\frac{U_s}{\sqrt{2\pi \alpha}} e^{i s k_F x}
e^{i(\theta - s \phi)}
\end{eqnarray}
with $s=1/-1$ for $R/L$ movers, the density
reads:
\begin{eqnarray}
\rho = \rho_0 -
\frac{\partial_x \phi}{\pi} + \frac{2}{2 \pi \alpha} {\rm Re}( e^{2 i k_F
x - 2 i \phi} )
\end{eqnarray}
where $\rho_0$ is the average density and $\alpha$
a UV cutoff. The (Tomonaga Luttinger) model $S_0 + S_{int}$
is equivalent, using $2 \pi \rho_s=-\partial_x \phi + s
\partial_x \theta$, to the quadratic action ${\cal S}_b[\phi]$
and to the CO model without disorder, with:
\begin{eqnarray}
&& K = T/T_g \label{KvsT}
\end{eqnarray}
hence any temperature can be reached by varying the Luttinger
parameter $K$, i.e. the interactions, with:
\begin{eqnarray}
&& K^2=(1+y_4-y_2)/(1+y_4+y_2) \\
&& y_i=g_i/(2 \pi v_F)
\end{eqnarray}
and the
sound velocity 
\begin{equation}
v=v_F \sqrt{(1+y_4)^2-y_2^2}
\end{equation}
which can be set to one by a
rescaling of space-time. Next, the (formal) perturbation of the
fermion model $S_f \to S_f + S_{dis}$ with 
\begin{eqnarray}
&& S_{dis}= - \frac{1}{\hbar} \int_{x \tau} W \rho \\
&& W(r)=\mu(r) + \tilde \xi(r) e^{2 i k_F x} + \tilde
\xi^*(r) e^{-2 i k_F x}
\end{eqnarray}
by a small (dynamic) random
potential $W(r)$ 
(neglecting for now higher harmonics) is
equivalent 
via ${\cal S}_b[\phi]=S_{CO}[u=2 \phi]$, to the disorder pertubation
in the CO model, i.e $h,\xi$ in (\ref{def_co}). The correspondance
reads $h^x=-\mu/(2 \pi \hbar)$, $h^\tau=0$,
$\frac{1}{T} \xi(r) = \tilde \xi(r)/(\pi \alpha \hbar )$. An
isotropic $h$ distribution is obtained by coupling a random field to
$\partial_\tau \phi$ but it makes no difference at large scale.
For $T<T_g$ small
disorder $g_0$ is relevant and grows under RG to finite value on the
fixed line \cite{tgpldrsb}. At a naive level (see below) the
relation (\ref{KvsT}) still holds since being unrenormalized (for
exact STS).

\subsection{Predictions from ''nearly conformal'' field theory}

The recent study \cite{ludwig} adresses a {\it free fermion} model in
presence of (isotropic) disorder $S_{ff} = S_0 +S'_{dis}$. We follow
the notations of Ref \cite{mudry} which uses disorder parameters
$g_A=\pi \sigma/T^2$ and $g_M \propto g_0$, shifted fermion fields
$\tilde \psi_s = e^{i(\theta - s \phi)}$ and the same
bosonic fields $\phi_{L/R}=\phi-s \theta$. One has chosen 
$\hbar=1=v_F$, $h_x=A_\tau/\pi$, $h_\tau=-A_x/\pi$. This model
was also found to be described by a line of FP with the
exact result (denoting $\zeta=1/(1+g_M/\pi)$
and $\phi=u/2$ the same bosonic phase as above):
\begin{eqnarray}
&& \partial_l g_M =0 \quad , \quad \partial_l g_A = \frac{(\zeta g_M)^2}{2 \pi^2} \label{result} \\
&& \langle \phi_a(z,\bar z) \phi_b(0) \rangle = - \frac{g_M^2
\zeta^4}{(2 \pi)^2} \ln^2 z \bar z -\zeta (\delta_{ab} + \frac{g_A
\zeta}{\pi} ) \ln z \bar z \nonumber
\end{eqnarray}
for real copies $a,b$ of the model (no replica). The vanishing beta
function for $g_M$ strongly suggests that the free fermion model is
exactly {\it on the CO fixed line}. From the above it means that the value of $K$
is shifted from $K=1$ (no disorder) down to $K=1/(1+g_M/\pi)$ (in presence of
disorder), to all orders in $g_M$ according to Ref.~\cite{ludwig,mudry}.
To lowest order in $g_M$ one easily finds such a shift
on model $S_0+S_{dis}$, e.g. considering the replicated action
\begin{eqnarray}
&& S_0[\psi_a]-\frac{1}{2 \pi^2} \int_{x \tau} (g_M O_M + g_A O_A) \\
&& O_M=\tilde \psi^\dagger_{Ra} \tilde \psi_{La} \tilde \psi^\dagger_{Lb}
\tilde \psi_{Rb}
\end{eqnarray}
While bosonization of the terms $a \neq b$
yields the $\cos(2(\phi_a-\phi_b))$ disorder terms
of the replicated CO action (see \ref{bare_action_2} below), the $a=b$ term
yields an 
interaction \cite{giam} as in Eq. (\ref{eq:fermions-periodic}) with $g_2=2
g_M$, compatible to lowest 
order with $K=1-g_2/\pi$. To understand higher orders one notes that the
interaction used in \cite{ludwig,mudry} is fully isotropic, and does
not correct $v$ (while $g_2$ corrects $v$ to order
$g_2^2$), hence involves also a $2 k_F$ coupling to current. Note that density only
couplings preserve $v K$ to all orders.
In bosonized form it reads 
\begin{equation}
\frac{g_M}{2 \pi^2} \int_{x \tau}
J_{aa} \bar J_{aa} = \frac{g_M}{2 \pi^2} \int_{x \tau} [ (\partial_\tau
\phi)^2 + (\partial_x  \phi)^2 ]
\end{equation}
where $J_{ab}=\psi_{Lb} \psi^+_{La}$.

\begin{figure}
\includegraphics[angle=-90,width=\linewidth]{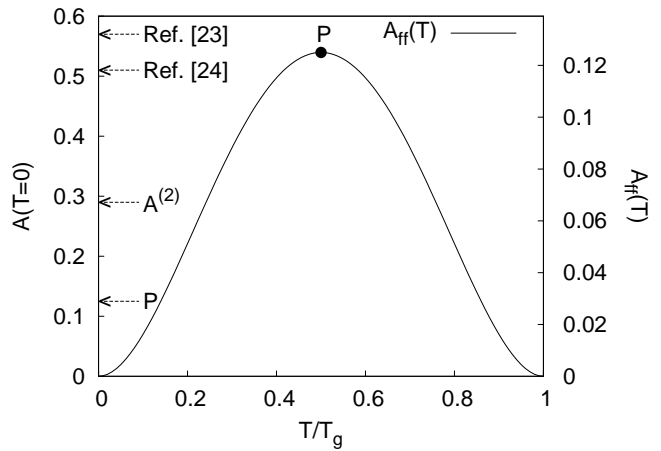}
\caption{{\bf Right $y$-axis}: $A_{ff}(T)$ given by formula (\ref{Aff}) obtained, as explained in
the text, from the theory of Ref. \cite{ludwig} and yielding
  $A_{ff}(T=0)=0$. {\bf Left $y$-axis}:
the exact ground state determination of disordered solid-on-solid models yield $A(T=0)
= a_2 = 0.57$ from Ref. \cite{rieger} and $A(T=0) =
2(2\pi)^2 B = 0.51$ from Ref. \cite{middleton}, shown together with the value
$A^{(2)}$ obtained by the present FRG method.}
\label{fig_1}
\end{figure}

The bosonised theory of Ref. \cite{ludwig} hence exhibits
superoughening, and is ''nearly conformal'' as the $h$ random field
(the $g_A$ sector) is a (Larkin type) random force trivially
eliminated by a shift using STS. The $\delta_{ab}$ term in
(\ref{result}) gives thermal connected correlations hence one
identifies $\zeta=T/T_g$. It implies the prediction (with
$\tau=1-T/T_g$):
\begin{eqnarray}
&& A_{ff}(T)=2 \tau^2 (1-\tau)^2 \label{Aff}
\end{eqnarray}
consistent with one loop at small $\tau$, exhibiting a maximum at
$\tau=1/2$ where $A_{ff}=1/8$ (see Fig. \ref{fig_1}) and {\it vanishing} at
zero 
temperature. Clearly $A_{ff}(T=0)=0$ is incompatible with numerical
results and equivalence of the models. The reentrant shape of the
$A_{ff}(T)$ is surprising as one expects stronger effect of disorder
at low $T$. This is reminiscent of reentrant phase boundaries found
in the random gauge XY model, due to neglecting operators (moments
of vortex fugacity) which become relevant at low $T$
\cite{nattermann95,dcpld}. Here higher harmonics of the $\exp(i
\phi)$ field are known to be generated as an interplay of
interactions and disorder. This can be seen in the
Haldane representation \cite{haldane}:
\begin{eqnarray}
&& \rho(r)=(\rho_0 - \frac{1}{\pi} \partial_x \phi(r)) \sum_p e^{2 i p
\rho_0 x} e^{- 2 i p \phi(r)}
\end{eqnarray}
which yields higher harmonics
$\cos(2p(\phi_a-\phi_b))$ in (\ref{bare_action_2}). Note that the
isotropic disorder $\int_{x\tau} \rho W + j V$ where $j$ is the
exact current, $\partial_\tau \rho+\partial_x j=0$, produces a
seemingly irrelevant coupling $\partial_x \phi(r) Re( e^{- 2 i
\phi(r)} \xi(r))$. Upon replicating its $aa$ contribution generates
a shift in $K$, and none in $v$. Whether the higher harmonics are generated in a
free fermion model (e.g. as in \cite{ludwig}) properly taking microscopic
cutoff into account deserves further 
investigations \cite{giam_prep}. Weak disorder counting shows them to be
irrelevant for  
$1/2<T/T_g<1$ but these dimensions could change along the fixed line
well before that. We now study the effect of higher harmonics using the
FRG.

\section{functional RG}

\subsection{Effective action and general structure of the RG flow}

We start by noting that $A_{ff}(T=0)=0$ is reminiscent of a {\it
dimensional reduction} result, i.e. an artefact of considering an
analytic disorder $R(u)$. This suggests a reexamination of the model
at low T using FRG. We now indeed show that the cusp in $R''(u)$ is
necessary to get $A(T=0) > 0$. Using replicas $u^a(r)=u^a_r$,
$a=1,.,n$ and averaging over $\xi,h$ we obtain the bare replicated
action (with $\overline {Z^n} = \int {\cal D}u e^{-S_0[u]}$):
\begin{eqnarray}
&&S_0[u] = \frac{H_{\rm rep}[u]}{T} =  \frac{\kappa}{2 T} \sum_a
\int_q u^a_q u^a_{-q} ({\cal G}^q_{l})^{-1} \label{bare_action_1} \\
&&- \frac{1}{2 T^2} \sum_{ab} \int_r \bigr[ R_0(u^{ab}_r) -\frac{1}{2} \nabla
u^{ab}_r \nabla u^{ab}_r G_0(u^{ab}_r) \bigr]
\label{bare_action_2}
\end{eqnarray}
with $u^{ab}_r=u^a_r-u^b_r$, $\int_r = \int d^d r$, $\int_q
\equiv \int d^dq/(2\pi)^d$. The bare propagator is
${\cal G}^q_{l} = q^{-2}c(q^2/2\Lambda_l^2,q^2/2\Lambda_0^2)$
where $\Lambda_0$ and $\Lambda_l = \Lambda_0 e^{-l}$ are UV and IR
cutoffs and $c(z,s)$ is a smooth cut-off function. We denote
$c(z,\infty)=1-c(z)$ with $c(0)=1$ and $c(\infty)=0$. At
the bare level, $R_0(u)=g_0 \cos(u)$ and $G_0(u) = \sigma$ in
(\ref{bare_action_2}). The effective action $\Gamma_l[u]$ (i.e. the Legendre tranform of the free energy in presence of sources) admits a
double expansion in number of replica (cumulants) and gradients, which starts
as
\begin{eqnarray}
&& \Gamma_l[u] =  \frac{\kappa}{2 T} \sum_a
\int_q u^a_q u^a_{-q} ({\cal G}^q_{l})^{-1} \\
&&- \frac{1}{2 T^2} \sum_{ab} \int_r \bigr[ R_l(u^{ab}_r) -\frac{1}{2} \nabla
u^{ab}_r \nabla u^{ab}_r G_l(u^{ab}_r) \bigr] + \cdots \nonumber
\label{bare_action_2}
\end{eqnarray}
as in
(\ref{bare_action_1},\ref{bare_action_2}) with (periodic)
renormalized function $R_l(u)$ and $G_l(u)$. Note that
the STS symmetry constrains the form of the one replica
term to be identical to the bare one. In $d=2$ a $p$-periodic $R_l$ becomes relevant
below $T=T_g^{(p)}=T_g 2 \pi/p$ and at $T=0$ is relevant below $d=4$. Higher
cumulants are irrelevant compared to $R_l$ near $d=4$. In $d=2$
a 1-loop approximation keeping only $R_l$
yields the exact $T_g$ \cite{chauve_creep_long}
and reasonable properties at low $T$ \cite{us_epl, schehr_simu}.
A specificity of $d=2$, not systematically studied with
the FRG (see however \cite{toner_radziho}) is that
the term $G_l$ must be retained. Indeed at $T=0$ and $d=2$ the whole function $G_l(u)$
is dimensionless at the bare level, and acquires anomalous dimensions due to
temperature and disorder. The $2$-point correlation function reads:
\begin{eqnarray}
\overline{\langle u_q u_{-q}\rangle} = \frac{T}{\kappa}{\cal G}^q_l +
\frac{({\cal G}^q_l)^2}{\kappa^2} (-R_l''(0) + G_l(0)q^2    )
\label{two_point_correl}
\end{eqnarray}
Near $T_g$, it was shown that $G_l(0) \sim l$ which yields,
setting $l = \log{(1/q)}$, the $\ln^2|r|$ behavior in $B(r)$
(\ref{roughness}). We now investigate how a similar behaviour,
$G_l(0) \propto l$ can arise also at low $T$ where, as for
$R_l(u)$, all harmonics in $G_l(u)$ must be kept.
We first set $\kappa=1$ in Eq. (\ref{bare_action_1}). Defining:
\begin{eqnarray} 
&&  \tilde T = S_d
\Lambda_l^{d-2} T \\
&& \tilde R_l(u) = S_d \Lambda_l^{d-4} R_l(u) \\
&& \tilde G(u) = S_d \Lambda_l^{d-2} G_l(u)
\end{eqnarray}
with $S_2 = (2\pi)^{-1}$,
the coupled RG equations for $\tilde R_l$ and $\tilde G_l$ have the
structure
\begin{eqnarray}
&&\partial_l \tilde T = (\epsilon -2) \tilde T \quad, \quad \partial_l
  \tilde R(u) = \beta_{\tilde R}(u) \nonumber \\
&&\partial_l \tilde
G(u) = (\epsilon -2) \tilde G(u) + \bar{\beta}_{\tilde G}(u)
\label{gen_struct}
\end{eqnarray}
with $\epsilon = 4-d$, $\beta_{\tilde R,\tilde G} \equiv \beta_{\tilde R,\tilde
  G}[\tilde R, \tilde {\sf
  R}'', \tilde R''', \tilde {\sf G}, \tilde G'... ]$ where $\tilde {\sf R}''(u)
= \tilde R''(u)-\tilde R''(0)$ and $\tilde {\sf G}(u) = \tilde G(u) -
\tilde G(0)$. Under the following assumptions : i) $\tilde G(u)$ is
continuous in $u =0$, ii) $\bar \beta_{\tilde G}(u)$ has a good limit
when $u \to 0^+$:
\begin{equation}
\bar \beta_{\tilde G}(0^+) = A_d
\end{equation} 
and iii) ${\sf
  G}(u)$ has a good fixed point for all $d$ one obtains that
$\tilde G(0) = {A_d}/{(2-\epsilon)}$, where $A_d$ admits an
$\epsilon$-expansion around $d=4$. Thus $\tilde G(0)$ is diverging
for $\epsilon \to 2$, yielding the announced behavior in $d=2$,
$\tilde G(0) = {A_2} l$. Note that this implies:
\begin{eqnarray}
&& \overline{\langle u_q \rangle \langle u_{-q} \rangle } \sim \frac{G(0)}{q^2}
\sim \frac{A_2}{S_2} \frac{ \ln(1/q) }{q^2}
\end{eqnarray}
hence, going back to real space $A(T=0)=A_2$. 

\subsection{Lowest order calculation}

We first compute, at $T=0$, $\beta_{\tilde R}, \beta_{\tilde G}$ to
${\cal O}({\tilde
  R^2})$. More details are given in Appendix \ref{appendixA}. From one loop 1PI graphs one finds
$\Gamma_l[u] = S_0[u] + \delta \Gamma[u]$:
\begin{eqnarray}
\hspace*{-0.5cm}\delta \Gamma[u] =  \frac{-1}{4 T^2} \sum_{ab}
  \int_{xy} ({\cal G}^{x-y}_{l})^2
  {\sf R}_0''(u_x^{ab}) {\sf R}_0''(u_y^{ab}) \label{delta_gamma}
\end{eqnarray}
excluding terms proportional to $T$ (tadpoles and some 3-replica
terms). Differentiating the local component of $\delta \Gamma[u]$ in
(\ref{delta_gamma}) w.r.t. the IR cutoff $\Lambda_l$ at fixed $R_0$,
and inverting $R_0 = R_l + {\cal
  O}(R_l^2)$, one obtains the standard $T=0$ 1-loop FRG
equation for $\tilde R_l$ in $d=4-\epsilon$:
\begin{eqnarray}
\partial_l \hat R_l(u) = \epsilon \hat R_l(u) + ({1}/{2}) ({\sf \hat
  R''}(u))^2 + {\cal O}(\tilde R{\sf \tilde G}, {\sf \tilde
    G}^2)\label{frg_g}
\label{frg_u}
\end{eqnarray}
with $\hat R_l(u) = 2 I_{00} \tilde R_l(u)$ where 
\begin{eqnarray}
I_{nm} =
S_d^{-1}\Lambda_l^{-d+4+n-m} \int_y y^n \partial_l {\cal G}^y_l
\nabla_y^m {\cal G}^y_l
\end{eqnarray} 
for $n,m$ even (see also Appendix \ref{appendixB}). Eq. (\ref{frg_u}) admits a
non-analytic, $p$-periodic fixed point solution \cite{bg_long} given
by 
\begin{eqnarray}
\hat R^*(pu) = (\epsilon p^4/(72))(1/36 - u^2(1-u)^2)
\end{eqnarray}
with $u
\in [0,1[$, yielding the amplitude of the cusp:
\begin{eqnarray}
\hat R^{*'''}(0^+) = - \hat R^{*'''}(0^-) = {p \epsilon}/{6} \label{r_tierce}
\end{eqnarray}

Next, the {\it bilocal} component of $\delta \Gamma[u]$ in
(\ref{delta_gamma}), when expanded in gradients generates a term
$\nabla u^{ab}_x \nabla u^{ab}_x G_l(u_x^{ab})$. Proceeding as above
one obtains \cite{ergus} (see Appendix \ref{appendixA}) up to ${\cal
  O}(\tilde R{\sf \tilde G}, {\sf \tilde
    G}^2)$:
\begin{equation}
\partial_l \tilde G_l(u) = (\epsilon-2)\tilde G(u) + ({I_{20}}/{d})
\tilde R'''(u)^2 \label{frg_g}
\end{equation}
Eq. (\ref{frg_g}) yields, in $d=2$, $\tilde G(0) \sim (I_{20}/2)
(\tilde R^{*'''}(0))^2 l$ which has a  non ambiguous value
(\ref{r_tierce}). It also leads to:
\begin{eqnarray}
{\sf
  \tilde G}(u) =  ({I_{20}}/{d})((\tilde R'''(u))^2 -(\tilde
R'''(0))^2)/(2-\epsilon)
\end{eqnarray}
thus satisfying the requirement iii)
above for $\epsilon < 2$. As shown below, the divergence for
$\epsilon = 2$ is cured by considering higher order terms in Eq.
(\ref{frg_g}). From (\ref{two_point_correl}),(\ref{frg_g}) a first
estimate $A^{(1)}$ of $A(T=0)$ in (\ref{roughness}) up to order
${\cal O}(\tilde R^2)$ is
\begin{eqnarray}
A^{(1)} = \frac{I_{20}}{2} (\tilde R^{*'''}(0^+))^2 = \frac{p^2
  \epsilon^2}{288} \frac{I_{20}}{I_{00}^2} \label{main_result}
\end{eqnarray}
As announced it is non zero at $T=0$ {\it only} because the fixed
point (\ref{r_tierce}) is {\it non-analytic}. The $p^2$ dependence
in Eq. (\ref{main_result}) yields that $A^{(1)}$ is independent of
$\kappa$. Indeed one has $B_{\kappa,p}(r) = \kappa^{-1}
B_{1,p/\sqrt{\kappa}}(r)$ if $B_{\kappa,p}(r)$ denotes the
correlation in (\ref{roughness}) for model (\ref{def_co}) with
$p$-periodic potential (and $h=0$). While $A_{d=2}$ is expected to be
universal, our truncation (\ref{main_result}) depends on the shape
of the IR regulator $c(z)$, with local sensitivity minimum around
the exponential, $c(x) = e^{-x}$ which yields $A^{(1)} = 0.881$.

\subsection{Higher order calculation} 

We now push the $T=0$ analysis further by computing $\beta_{\tilde
  R}, \beta_{\tilde G}$ up to order ${\cal O}(RG, G^2)$, which are of
order ${\cal O}(1)$ in $d=2$. From $\delta \Gamma[u]$ to one loop, 
one finds:
\begin{eqnarray}
&&\beta_{\tilde R}^{T=0} = \epsilon \tilde R + I_{00} {\sf \tilde R}''^2 +
2I_{02} {\sf \tilde R''}{\sf \tilde G} + I_{04} {\sf \tilde G}^2
... \label{frg_r_2}\\
&&\bar \beta_{\tilde G}^{T=0} = ({I_{20}}/{d}) \tilde R'''^2
+ \gamma_1 \tilde R''' \tilde
G' +  2I_{00}
{\sf \tilde R''}\tilde G''
 \label{frg_g_2} \\
&&+ 2I_{02} {\sf \tilde G}\tilde G'' +
\gamma_2  \tilde G'^2 + ... \nonumber
\end{eqnarray}
where  (see
  Appendix \ref{appendixB} for more details):
\begin{eqnarray}
&& \gamma_1 = 4(I_{00} + {K_{0}}/{d} + {I_{22}}/{2d}) \\
&& \gamma_2 = 6I_{02} + {I_{24}}/{d} + {K_2}/{d} \\
&&  S_d^{-1}
\Lambda_l^{-d+4-n} \int_y y_i \partial_l {\cal G}^y_l \nabla_j
\nabla^n {\cal G}^y_l = \delta_{ij}K_n/d 
\end{eqnarray}
Remarkably,
(\ref{frg_r_2}, \ref{frg_g_2}) admit a simple solution:
\begin{eqnarray}
\tilde {\sf R}''(pu) = p^2 a_d u(1-u) , \quad  \tilde {\sf G}(pu) =
p^2 \alpha a_d u (1-u)  \label{sol_rg}
\end{eqnarray}
with:
\begin{eqnarray}
a_d = \frac{\epsilon}{12(I_{00} + 2 \alpha I_{02} + \alpha^2 I_{04} )}
\label{eq_ad} 
\end{eqnarray}
and $\epsilon=4-d$. Eqs. (\ref{frg_g_2}), (\ref{sol_rg}) satisfy all above
requirements i), ii) iii). It yields a value for $A_d$, up to order ${\cal
O}(\tilde R{\sf
  \tilde G},{\sf \tilde G}^2)$ at one loop, for a $p$-periodic
  potential:
\begin{eqnarray}\label{A2}
A_d = \frac{p^2 \epsilon^2}{144} \frac{I_{20}/d + \alpha
  \gamma_1 + \alpha^2 \gamma_2 }{(I_{00} +
  2\alpha I_{02} + \alpha^2 I_{04})^2}
\end{eqnarray}
yielding back (\ref{main_result}) if $\alpha$ is (formally) set to
zero. Here $\alpha$ is the solution of a cubic
equation for $\epsilon < 2$ (with $\alpha = {\cal O}(\epsilon)$ near
$d=4$), and of a quadratic equation in $d=2$ 
\begin{eqnarray}
\alpha^2 (I_{02} + \gamma_2)
+ \alpha (I_{00}+\gamma_1) 
+ \frac{I_{20}}{2} = 0 \label{Eq_alpha}
\end{eqnarray} 
Among the two solutions of Eq. (\ref{Eq_alpha}), only one is 
physical, and yields $\tilde G(0) \propto
A_2 l$ and an estimate $A^{(2)}$ of $A(T=0)$ up to order ${\cal
O}(\tilde R{\sf
  \tilde G},{\sf \tilde G}^2)$ at one loop. 
Using relations valid in $d=2$ it can also be written as:
\begin{eqnarray}
A^{(2)}
= -\frac{p^2 \epsilon^2 \alpha}{144} \frac{\alpha I_{02} + I_{00}}{(I_{00} +
  2\alpha I_{02} + \alpha^2 I_{04})^2}
\end{eqnarray}
$A^{(2)}$ is independent of $\kappa$ and rather stable as the
IR cutoff $c(z)$ is varied. A Gaussian $c'(x) \propto
e^{-(x-1/2)^2/(2\sigma^2)}$ reaches a local maximum with $\sigma =
1/4$ and $A^{(2)} = 0.29$. Thus the numerical results, $A(T=0) = a_2
= 0.57$ of Ref. \cite{rieger} (notice that $a_2$ refers here to the notation
used in Ref. \cite{rieger} and is different from $a_{d=2}$ as in our
Eq. (\ref{eq_ad})) and $A(T=0) = 2(2\pi)^2 B = 0.51$ of 
Ref. \cite{middleton} seem to lie in between our estimates
$A^{(1)},A^{(2)}$ (see Fig. \ref{fig_1}).

%
%
%

\subsection{Extension to low temperature}

To be convincing the scenario should be stable to a small
temperature $T$. To one loop (\ref{gen_struct}) become:
\begin{eqnarray}
&&\partial_l \tilde R(u) =  \tilde T \tilde R''(u) - J_2 \tilde T
  {\sf \tilde G}(u) + \beta_{\tilde R}^{T=0}(u) \\
&&\partial_l \tilde G(u) = (\epsilon -2) \tilde G(u) + \tilde T
\tilde G''(u) +
  \bar{\beta}^{T=0}_{\tilde G}(u)
\label{gen_struct_T}
\end{eqnarray}
where $J_2 = (S_d)^{-1}\Lambda_l^{d-4}\partial_l \nabla^2 {\cal
  G}^{y=0}_l$. At low $T$, and dimension $d=2$, we look for a solution of
  Eq. (\ref{gen_struct_T}) 
  where for $u \gg T$ the functions $\tilde R$ and $\tilde G$ are given
by the $T=0$ solution (\ref{sol_rg}), while for small $u \sim T$
there is a thermal boundary layer (TBL) solution of the form 
\begin{eqnarray}
{\sf \tilde R''}(u) = T r''(u/T) \quad, \quad \tilde G(u) = A_2 l  + c + T
g(u/T) 
\end{eqnarray}
where, with no loss of generality, $g(0)=r''(0)=0$. The functions $r''(z)$ and
$g(z)$ satisfy the coupled equations (for $p=1$)
\begin{eqnarray}
0 &=& -a_2 \frac{z^2}{6} + r'' - J_2 g + I_{00} r''^2 +
2I_{02} r''g + I_{04} g^2 \nonumber \\
A_2 &=& g'' + \frac{I_{20}}{d} r'''^2 + \gamma_1 r'''g' + 2I_{00} r''g''
\nonumber \\  
&+& 2I_{02} gg'' + \gamma_2 g'^2 \label{tbl} 
\end{eqnarray}
where we have used $\tilde R''(0) = -a_d/6 p^2$
and the notations $r \equiv r(z), g \equiv g(z)$. From Eq. (\ref{tbl}), one
easily checks the large $|z|$ behaviors
\begin{eqnarray}
&& r''(z) \sim a_d |z| + r_0 + O(|z|^{-1}) )\\
&&  g(z) \sim \alpha a_d |z| + g_0 + O(|z|^{-1}) ) \label{matching} 
\end{eqnarray}
which match correctly the cusp (\ref{sol_rg}) outside the TBL. 
The constants are $r_0=(J_2 I_{02} + I_{04})/(2 I_{02}^2 - 2 I_{04} I_{00})$
and $g_0=-(1+2 r_0 I_{00})/(2 I_{02})$.

\begin{figure}
\includegraphics[angle=0,width=\linewidth]{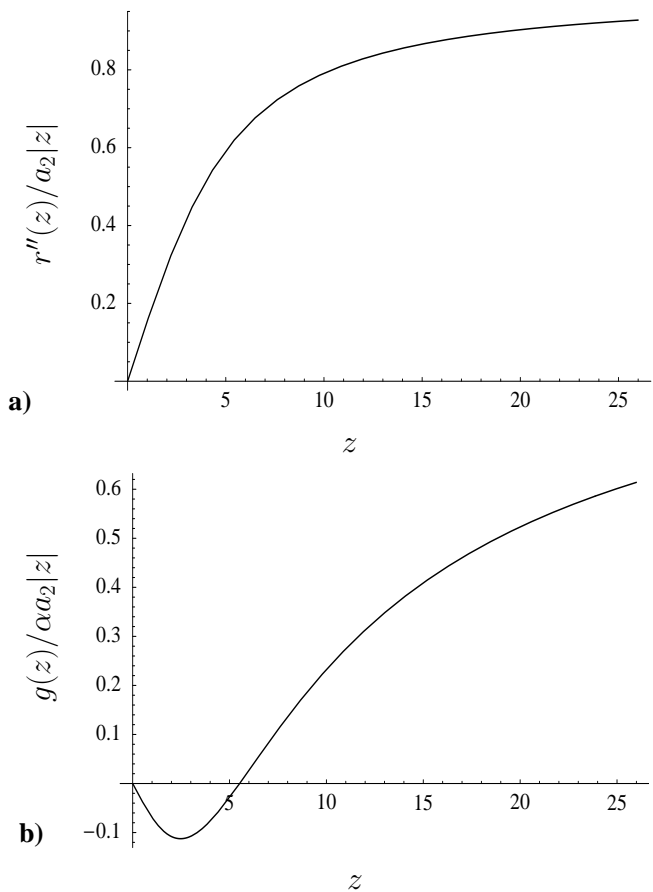}
\caption{Numerical solution of the coupled RG equations (\ref{tbl}) in
  dimension $d=2$, for $p=1$. We have used an
  exponential cutoff function $c(x) = e^{-a x}$. {\bf a)} : Plot of
  $r''(z)/a_2 |z|$ as a function of 
  $z >0$. {\bf b)} Plot of $g(z)/\alpha a_2 |z|$ as a function of
  $z$. Notice that the convergence towards the asymptotic behavior
  (\ref{matching}) is quite slow. We have checked that both curves converge
to unity at very large $|z|$.}\label{tbl_num}  
\end{figure}

We have solved numerically these RG equations (\ref{tbl}) and
checked that it admits a well defined solution~: this is depicted in
Fig. \ref{tbl_num} where we plot $r''(z)/a_2 |z|$ (respectively $g(z)/\alpha
a_2 |z|$) as a function of $z$ in Fig.\ref{tbl_num} a) (respectively
Fig.\ref{tbl_num} b)).    
Hence this shows that the scenario is consistent: in
fact these one loop equations, if naively continued already predict
a $\ln^2$ behavior (\ref{roughness}) at all $T$ and the correct
$T_g$, with, near $T_g$, $\tilde R(u) \sim \cos{u}$ since only the
lowest harmonic is relevant. A more complete interpolation of $A(T)$ deserves
  further investigations. 

\section{conclusion}

To conclude we applied the Functional RG to the random phase
Sine-Gordon model, when gradient expansion is important. We showed
that the super-roughening amplitude $A(T)$ does not vanish at $T=0$
thanks to the non-analyticity of the fixed point function $R(u)$,
which avoids dimensional reduction. The one loop estimates give an
order of magnitude consistent with numerics. Our study suggests that
either the mapping between free fermions model and the CO model fails
below some temperature or that new operators become relevant, e.g.
higher harmonics, as is the case for interacting fermions models.
Such a scenario was demonstrated in random field spin models
\cite{feldman_rf}. Whether the path integral transformations of Ref
\cite{ludwig} hold at low $T$ needs to be clarified. A procedure
exists \cite{pierre_rofu} to measure the FRG functions described
here, in numerics and check the scenario. We hope that it allows
progress in issues common to fermion models, localization and
periodic pinned systems.

\acknowledgments
We thank N. Andrei, T. Giamarchi, A. Ludwig and E. Orignac for discussions.
GS acknowledges support from the EC HPRN-CT-2002-00307, DYGLAGEMEM
and PLD from ANR program 05-BLAN-0099-01.
\vskip-0.5cm

\vspace*{1cm}

\appendix

\section{gradient expansion}
\label{appendixA}

The calculation of of the
effective action up to order $R_0^2$ yields:
\begin{eqnarray}
&&\delta \Gamma[u] =  - \frac{1}{2 T^2} [ \sum_{ab} \int_x \langle
R_0(u_x^{ab} + 
\delta u_x^{ab}) \rangle \\
&&+ \frac{1}{4 T^2} \sum_{abcd} \int_{xy} \langle R_0(u_x^{ab} + \delta
u_x^{ab}) 
R_0(u_y^{ab} + \delta u_y^{ab}) \rangle] \nonumber
\end{eqnarray}
where $\langle...\rangle$ means average over the $\delta u$ with the free
propagator, keeping only the 
1PI graphs. This gives:
\begin{widetext}
\begin{eqnarray}
&&\delta \Gamma[u] =  - \frac{1}{2 T^2} [ \sum_{ab} \int_x R_0''(u_x^{ab})
{\cal G}_l^{0}  + \int_{xy} ({\cal G}_l^{x-y})^2 [ \sum_{ab}  \frac{1}{2}
{\sf R}_0''(u_x^{ab}) {\sf R}_0''(u_y^{ab}) + \sum_{abc} R_0''(u_x^{ab})
R_0''(u_y^{ac}) ]  
\end{eqnarray}
\end{widetext}
where ${\sf R}_0''(u)=R_0''(u) - R_0''(0)$. 
The last term is a third cumulant proportional to temperature, and here we
ignore it. 
We now expand the second term in gradients. Consider a general bilocal
functional:
\begin{eqnarray}
\int_{xy} F(u_x,u_y,y-x) 
\end{eqnarray}
We can either use $z=(x+y)/2$, $t=y-x$ and expand everything to second order
in $t$ :
\begin{widetext}
\begin{eqnarray}
&& \int_{zt} F(u_z + \frac{1}{2} (t \cdot \nabla) u_z + \frac{1}{8} (t \cdot \nabla)^2 u_z,
u_z - \frac{1}{2} (t \cdot \nabla) u_z + \frac{1}{8} (t \cdot \nabla)^2 u_z,t)
\\ 
&& = 
\int_{zt}  \frac{1}{8} (t \cdot \nabla) u_z (t \cdot \nabla) 
u_z (\partial_1 - \partial_2)^2 F(u_z,u_z,t) 
+ \frac{1}{8} (t \cdot \nabla) (t \cdot \nabla) u_z (\partial_1 + \partial_2) F(u_z,u_z,t)
\end{eqnarray}
\end{widetext}
Using also that $\int_{t} F(u,u,t)=0$ (we assume that the local part has already been substracted) and performing integration by part to get rid of the $\nabla \nabla u F(u,u,t)$
term, we obtain a term containing $(\partial_1 - \partial_2)^2  - (\partial_1 + \partial_2)^2 =
- 4 \partial_1 \partial_2$ and thus finally one obtains the gradient
expansion as:
\begin{eqnarray}
&&\int_{xy} F(u_x,u_y,y-x)  \nonumber \\
&&= - \int_{zt} \frac{1}{2} (t \cdot \nabla) u_z (t \cdot \nabla) u_z \partial_1 \partial_2 F(u_z,u_z,t)
\end{eqnarray}
Alternatively one can also perform the expansion as
\begin{eqnarray}
&&\int_{xt} F(u_x,u_y,y-x) \\ 
&&= \int_{xy} F(u_x,u_x + (t \cdot \nabla) u_x + \frac{1}{2} (t \cdot
\nabla)^2 u_x,t) \nonumber \\ 
&&= \int_{xt} \frac{1}{2} [ (t \cdot \nabla) u_x (t \cdot \nabla) 
u_x \partial_2^2 \\ \nonumber 
&&+  (t \cdot \nabla) (t \cdot \nabla) u_x \partial_2 ]
F(u_x,u_x,t) 
\end{eqnarray}
and integration by parts then yields the same result as above. Thus one finds that the part
of interest of the effective action is:
\begin{eqnarray}
&& \delta \Gamma[u] = \frac{1}{8 T^2} \sum_{ab} \int_{x} R_0'''(u_x^{ab})
R_0'''(u_x^{ab}) \nonumber \\
&& \times \int_{t} ({\cal G}_{t}^2 - \delta(t) \int_{s} {\cal G}_{s}^2) (t \cdot \nabla) u^{ab}_x (t \cdot \nabla) u^{ab}_x
\end{eqnarray}
yielding the result given in the text.

\section{calculation of integrals}
\label{appendixB}

In this appendix, we give the detailed calculation of the coefficients $I_{n,m}, K_n$ entering the RG equations given in the text. One 
introduces the notation ${\cal G}_{l}^y = \Lambda_l ^{d-2} \gamma_l^{\Lambda_l
  y}$. Focusing on the large $l$ limit, one defines $\gamma_{l \to \infty}^y =
g^y$ and $\partial
g^y$ as:
\begin{equation}
g^y = \int_q e^{iqy} \frac{1}{q^2}(1-c(q^2/2)) \quad, \quad \partial
g^y = -\int_y c'(q^2/2) e^{iqy}
\end{equation}
Let us start with $I_{00}$: 
\begin{eqnarray}
I_{00} &=& \frac{1}{S_d} \int_y \partial g^y g^y = - \frac{1}{S_d}
\int_q \frac{c'(q^2/2)}{q^2} (1-c(q^2/2)) \nonumber \\
&=& -\frac{1}{2} \int_0^\infty du \frac{c'(u)}{u} (1-c(u)) \quad
\rm{in} \quad d=2  
\end{eqnarray}
In a similar way, one computes the coefficients (in $d=2$)
\begin{eqnarray}
I_{02} &=&  -\frac{1}{2} \\
I_{04} &=& -2 \int_0^\infty du u c'(u)(1-c(u)) 
\end{eqnarray}
Next, we compute $I_{20}$
\begin{eqnarray}
I_{20} &=& \frac{1}{S_d} \int_y y^2 \partial g^y g^y  \\
&=&  \frac{2}{S_d}
\int_q \frac{1}{q^2}c''(q^2/2) (1-c(q^2/2)) \nonumber \\
&& + \frac{1}{S_d}\int_q
c'''(q^2/2)(1-c(q^2/2))    \nonumber \\
&=& \int_0^\infty du (1-c(u)) (c''(u)/u+c'''(u))  \quad
\rm{in} \quad d=2    \nonumber
\end{eqnarray}
Similarly, one has (in $d=2$)
\begin{eqnarray}
I_{22} &=& -2\int_0^\infty du (1-c(u))(c''(u) + uc'''(u)) \\
I_{24} &=& 4\int_0^\infty du (1-c(u)) (uc''(u) + u^2c'''(u))
\end{eqnarray}
Finally one has
\begin{eqnarray}
K_{0} &=& \frac{1}{S_d} \int_y \partial g^y (\nabla^j g^y) y^j \nonumber \\
&=& -
\frac{1}{S_d} \int_q c''(q^2/2)  (1-c(q^2/2)) \\
&=& - \int_0^\infty du c''(u) (1-c(u)))  \quad
{\rm in} \quad d=2 
\end{eqnarray}
and similarly, in $d=2$
\begin{equation}
K_{2} = 8 \int_0^\infty du u c''(u) (1-c(u))
\end{equation}

\begin{thebibliography}{50}
\bibitem{ergus}
We checked that the exact RG on the effective action ~\cite{us_erg}
yields the same equations (\ref{frg_u}, \ref{frg_g}).
\end{thebibliography}
\end{document}